\documentclass[12pt]{article}
\usepackage{geometry}
\usepackage{a4}
\usepackage{graphicx}
\usepackage{epsf}
\usepackage{amsmath}
\usepackage{amssymb}
\usepackage{cite}
\newcommand{\be}{\begin{equation}}
\newcommand{\ee}{\end{equation}}

\newcommand{\Rmnum}[1]{\expandafter\@slowromancap\romannumeral #1@}
\newcommand{\bea}{\begin{eqnarray}}
\newcommand{\eea}{\end{eqnarray}}

\begin{document}
\def\A{{\mathbb{A}}}
\def\B{{\mathbb{B}}}
\def\C{{\mathbb{C}}}
\def\R{{\mathbb{R}}}
\def\s{{\mathbb{S}}}
\def\T{{\mathbb{T}}}
\def\Z{{\mathbb{Z}}}
\def\W{{\mathbb{W}}}
\begin{titlepage}
\title{Information Geometry and Quantum Phase Transitions in the Dicke Model}
\author{}
\date{
Anshuman Dey, Subhash Mahapatra, Pratim Roy, \\ Tapobrata Sarkar
\thanks{\noindent E-mail:~ deyanshu, subhmaha, proy, tapo @iitk.ac.in}
\vskip0.4cm
{\sl Department of Physics, \\
Indian Institute of Technology,\\
Kanpur 208016, \\
India}}
\maketitle
\abstract{
\noindent
We study information geometry of the Dicke model, in the thermodynamic limit. The scalar curvature $R$ of the Riemannian metric tensor
induced on the parameter space of the model is calculated. We analyze this both with and without the rotating wave
approximation, and show that the parameter manifold is smooth even at the phase transition, and that the scalar curvature is continuous across the phase boundary.}
\end{titlepage}

\section{Introduction}

The physics of phase transitions has been the focus of intense research over the last century. Information theoretic analyses of the same are
relatively recent, and have mostly found applications in relation to second order classical phase transitions (CPTs) \cite{brodyhook}. The essential idea herein 
is geometric in nature \cite{rupp1} and rests on the definition of a Riemannian metric on the equilibrium thermodynamic state space of a system. 
Ruppeiner \cite{rupp} has conjectured that the scalar
curvature $R$ arising out of such a metric is related to the correlation volume of the system, i.e $|R|\sim \xi^d$, where $d$ is the system dimension. This association
follows from thermal fluctuation theory, with the understanding being that the greater is the ``distance'' between two equilibrium thermodynamic states, the
lesser is the probability that these are related by a fluctuation (for a nice, pedagogical explanation, see \cite{ruppajp}). 
The relation between the scalar curvature and the correlation volume has been tested
for a variety of classical models \cite{rupp} and has also been applied to the theory of first order phase transitions in \cite{rsss}, 
and show excellent agreement with experimental results for liquid-gas systems. (For related recent results in liquid-liquid phase transitions and for Lennard-Jones
fluids, see \cite{tapo2}, and \cite{may}). 
It is by now established that for classical systems, $R$ diverges on the spinodal curve, and its equality in co-existing phases indicates a first 
order phase transition. These two aspects can be combined to map the phase diagram of the system, purely from a geometric perspective.

Information theoretic methods can be applied to quantum mechanical systems as well. The pioneering work in this direction was carried out
by Provost and Vallee \cite{pv}, who defined a Riemannian metric tensor from the Hilbert space structure of quantum states of a system. The application of this method
to the study of zero temperature second order quantum phase transitions (QPTs) is more recent, and appeared (following the results of \cite{zan2}) 
in the seminal work of Zanardi et. al. \cite{zan1}. In particular, these authors showed
that the scalar curvature arising out of the metric on the manifold of coupling constants\footnote{This will be referred to as the parameter manifold, or the 
information geometric manifold.} in the transverse XY spin chain is an indicator of QPTs in the system. 

Let us remind the reader that information geometry of QPTs has very different qualitative features compared to that of CPTs. 
This is not unexpected, given that the physical mechanisms involved in the two situations are different. 
For classical systems, the geometric parameters are thermodynamic variables and their Legendre transforms (hence information geometry in the context of CPTs is also referred to 
as thermodynamic geometry), whereas for quantum systems, the geometry is characterized by parameters that appear in the Hamiltonian, i.e the coupling constants of
the theory. The relation between the scalar curvature and the correlation length ($|R|\sim \xi^d$) that holds for CPTs, does not, in 
general, hold for QPTs, as we will show. 

The work of \cite{zan1}
puts information theoretic studies of QPTs into perspective, along with the more popular methods involving fidelity, defined as the modulus of the overlap between
two infinitesimally separated ground state wave functions \cite{zan2}, and fidelity susceptibility (see, e.g. \cite{amit1}. For an excellent review, see \cite{gu}). 
However, although the literature on the fidelity approach to the study of QPTs is vast, less attention has been paid to the geometric structure of the parameter manifold, 
and in particular, its scalar curvature $R$. While for classical spin models and fluid systems, $R$ always diverges at a second order phase
transition, the same is not true for QPTs. For example, it was shown in \cite{zan1} that for the transverse XY spin chain, $R$ develops a discontinuity at 
the Ising transition lines and does not diverge there. While an argument arising out of first order perturbation theory shows that the components of the metric tensor 
on the parameter manifold generally diverge at a QPT \cite{zan1} (this may not be true in some specific examples, see \cite{zan1a}), the same is thus 
not true for the scalar curvature. 
The latter is however a useful quantity to study, since it is invariant under coordinate transformations (of the parameter space), 
as opposed to the metric components (that transform as rank two tensors under such transformations).
For two dimensional parameter manifolds 
which is the case we will be concerned with in this paper, standard textbook results 
indicate that the Riemann curvature tensor \cite{blau} has only
one independent component, and hence leads to a unique scalar curvature.  Whereas divergences in the 
components of the metric tensor might be due to specific coordinate choices, and can be removed by (possibly singular) coordinate transformations,
those of the scalar curvature correspond to ones that cannot be removed by such transformations, and therefore characterize the global (i.e, coordinate independent) 
properties of the manifold. 

It is important and interesting to study the behavior of the 
parameter manifolds, for models exhibiting QPTs, via its scalar curvature, as this is a coordinate free approach and is further expected to give insights into
the difference in the geometrical nature of QPTs vis a vis CPTs. To the best of our knowledge, this has not been done before. 
The main purpose of this paper is to study the behavior of $R$ for 
the Dicke model of quantum optics, in the thermodynamic limit (TDL), for finite values of de-tuning. 
We show that in this model, although the components of the metric tensor diverge at QPT, 
$R$ is regular at, and continuous across, the phase boundary, and hence the parameter manifold is smooth, even at the location of the phase transition. 
Our results indicate that the global behavior of the parameter manifold varies widely between models, and does not allow for an universal characterization 
of QPTs, a situation that is markedly different from the classical case. 

This paper is organized as follows. In the next section, we briefly review information geometry and
the results of \cite{zan1} for the transverse field XY model. This section is mostly meant to set the notations and conventions for what is presented in 
the rest of the paper, but towards the end, we indicate some novel features of $R$. 
In section 3, we analyze the Dicke model of quantum optics, and calculate the associated scalar curvature, and the fidelity susceptibility. This is 
done both with and without the rotating wave approximation. Section 4 ends with 
our conclusions and directions for future study, where we also indicate some preliminary results on similar analyses for the 
Lipkin-Meshkov-Glick model and the Kitaev honeycomb model. 

\section{Information Geometry and the Transverse XY Model}

In this section, we briefly review the basics of information geometry, and the results of \cite{zan1} in the context of the transverse field XY model. The essential idea here
\cite{pv} is to use two infinitesimally separated quantum states, and construct (upto second order) the quantity
\begin{equation}
|\psi\left(q +dq\right) - \psi\left(q\right)|^2 = \langle \partial_i \psi|\partial_j \psi\rangle dq^idq^j = \alpha_{ij}dq^idq^j
\label{pv1}
\end{equation}
where $q^i$ (collectively denoted as $q$ in the l.h.s of eq.(\ref{pv1})) denotes the parameters on which the wave function depends on, and
$\partial_i$ is a derivative with respect to $q^i$. The $\alpha_{ij}$s are not gauge invariant, and 
in \cite{pv}, it was shown that a meaningful second rank tensor can be defined as
\begin{equation}
g_{ij} = \alpha_{ij} - \beta_i\beta_j;~~~~\beta_i = -i\langle\psi\left(q\right)|\partial_i\psi\left(q\right)\rangle
\label{metric}
\end{equation}
where the quantities $g_{ij}$ specify a metric on the parameter manifold, induced from the 
natural structure of the Hilbert space of quantum states. This metric can be shown to be positive definite (i.e the diagonal elements and the principal minors are
positive definite), and hence one obtains a Riemannian structure 
via a distance, i.e a line element $ds^2 = g_{ij}dq^idq^j$ which can be used to study the system under a variation of the parameters $q^i$. 
Note that the physical meaning of this metric (and the line element) lies in the fact that the larger the distance, the higher is the statistical distinguishability of the two states. 
For the purpose of studying quantum phase transitions, it is natural to focus on the ground state overlap and the corresponding induced metric \cite{zan2},\cite{zan1}. 
For the case of a two-parameter model, i.e $q^i = \left(q^1, q^2\right)$, the scalar curvature for the manifold with this metric can be computed 
via the standard formula 
$R = -\frac{1}{\sqrt{g}}\left({\mathcal A} + {\mathcal B}\right)$, where 
\begin{eqnarray}
{\mathcal A} &=&  \frac{\partial}{\partial q^1}\left[\frac{g_{12}}{g_{11}\sqrt{g}} \frac{\partial g_{11}}{\partial q^2} - \frac{1}{\sqrt{g}}\frac{\partial g_{22}}{\partial q^1}\right]\nonumber\\
{\mathcal B} &=& \frac{\partial}{\partial q^2}\left[\frac{2}{\sqrt{g}} \frac{\partial g_{12}}{\partial q^1} - \frac{1}{\sqrt{g}}\frac{\partial g_{11}}{\partial q^2} -
\frac{g_{12}}{g_{11}\sqrt{g}}\frac{\partial g_{11}}{\partial q^1}\right]
\label{scalarcurvature}
\end{eqnarray}
Here, $g_{ij},~i,j=1,2$ are a function of the the parameters $\left(q^1,q^2\right)$, and $g$ is the determinant of the metric of eq.(\ref{metric}). 

One of the few models that yield analytic results for the scalar curvature is the transverse XY model \cite{zan1}, which we briefly review. The Hamiltonian 
for the model with $(2N + 1)$ spins is
\begin{equation}
H_{\rm XY} = -\left[\sum_{j = -N}^{N}\frac{1 + \gamma}{4}\sigma_j^x\sigma_{j+1}^x + \frac{1-\gamma}{4}\sigma_j^y\sigma_{j+1}^y - \frac{h}{2}\sigma_j^z\right]
\end{equation}
where the $\sigma^i$, $i=x,y,z$ are Pauli matrices, $\gamma$ is an anisotropy parameter and $h$ is the magnetic field. 
Diagonalizing this Hamiltonian with the Jordan-Wigner, Fourier and Bogoliubov transformations, 
it can be checked \cite{zan1} that with the coordinates $\left(q^1,q^2\right) = \left(h,\gamma\right)$, the components 
of the metric induced on the parameter manifold is given by $g_{ij} = \frac{1}{4}\sum_{k>0} \frac{\tilde{g}_{ij}}{\Delta_k}$, where 
$\Delta_k = \left((h-{\rm Cos} (x_k))^2+\gamma ^2 {\rm Sin} ^2(x_k)\right)^2$ and 
\begin{equation}
{\tilde g}_{hh} = \gamma^2{\rm Sin}^2 (x_k);~~
{\tilde g}_{\gamma\gamma} = {\rm Sin} ^2(x_k) \left[h-{\rm Cos}(x_k)\right]^2;~~
{\tilde g}_{h\gamma} = \gamma  {\rm Sin}^2(x_k)\left[{\rm Cos} (x_k)-h\right]
\end{equation}
where $x_k = \frac{2\pi k}{L}$, $L = 2N + 1$. In the TDL, the sum over $k$ can be replaced by an integral, which can be evaluated 
in the complex plane \cite{zan1}, and it can be shown that metric components have four poles (only $g_{\gamma\gamma}$ has
an additional pole at the origin), located on the complex plane at values of the coordinates 
\begin{equation}
\lambda^{\pm}_1 = \frac{h \pm \sqrt{h^2 + \gamma^2 - 1}}{1-\gamma};~~\lambda^{\pm}_2 = \frac{h \pm \sqrt{h^2 + \gamma^2 - 1}}{1+\gamma}
\end{equation}
Evaluating the integrals then imply that some of the metric components are divergent at $\gamma=0$ and at $h = \pm 1$ (see below), and the scalar
curvature is
\begin{eqnarray}
R &=& -\left(\frac{32}{L}\right) \frac{1}{1 - \lambda_2^+\lambda_2^-};~~~|h| <1,~\gamma >0\nonumber\\
~ &=& \left(\frac{32}{L}\right)\frac{1}{1- \lambda_1^-\lambda_2^-};~~~|h|>1,~\gamma>0
\label{Rxy}
\end{eqnarray}
with similar expressions for $\gamma<0$. This is equivalent to the result derived in \cite{zan1} (as can be checked by substituting the 
values of $\lambda_{1,2}^{\pm}$ in eq.(\ref{Rxy})), but the usefulness of 
eq.(\ref{Rxy}) is that, using the results of \cite{barouch},\cite{bunder}, it reveals that
for the region $|h|<1$, $|R| \propto \left(1 - e^{-1/\xi}\right)^{-1}$, 
where $\xi$ is the correlation length inside the circle $h^2 + \gamma^2 = 1$. No such relation exists for the region $|h| >1$. Hence, on the 
line $\gamma = 0$ (with $|h| <1$), the scalar curvature diverges, as does the correlation length, but this is not the case for the Ising transition lines. 
This indicates that, as alluded to in the introduction, there is no simple relation between the scalar curvature and the correlation length at a second order quantum
phase transition, as opposed to classical cases. 

We also emphasize that in this example, the scalar curvature $R$ does {\it not} show any special behavior near the Ising critical lines. It does show a
discontinuity along these lines \cite{zan1}, but no geometrical meaning can be attributed to it, since the information metrics are different in the 
regions separated by these lines. The only genuine (i.e coordinate independent) curvature singularity is on the line $\gamma = 0$ for $|h| <1$, where $R$ diverges, 
and is related here to the correlation length, as we have mentioned. The singularities of the metric near the Ising transition lines are thus simply artifacts of 
coordinate choices. For example, in the region $|h|<1$, the metric can be written in two equivalent forms,
\begin{equation}
ds^2 = \frac{L}{16\gamma}\left(\frac{dh^2}{1-h^2} + \frac{d\gamma^2}{\left(1+\gamma\right)^2}\right) \equiv  
\frac{L}{16\gamma}\left(d{\tilde h}^2  + \frac{d\gamma^2}{\left(1+\gamma\right)^2}\right)
\label{diffmetrics}
\end{equation}
by a redefinition of coordinates ${\tilde h} = {\rm Sin}^{-1}\left(h\right)$. Whereas the first metric of eq.(\ref{diffmetrics}) has a divergence in the limit $|h| \to 1$ in its first component, the 
second metric does not have any divergence (apart from that at $\gamma = 0$ which is a genuine singularity and cannot be removed by a coordinate re-definition). 
The coordinate transformation indicated above is undefined at $h=\pm 1$, but note that in the region $|h| < 1$,
it gives rise to the same scalar curvature as does the first metric of eq.(\ref{diffmetrics}), and thus describes globally the same manifold as the first metric. 
For the region $|h| >1$, the metric can be calculated in a similar fashion,
and it turns out that $R$ does not show any divergence in this region as well, except at the multi critical point (MCP), 
$\gamma = 0, |h| \to1$. The information geometric manifold, in this region, is thus smooth (excepting for the MCP), and the divergences of the metric tensor 
can again be removed by appropriate coordinate transformations.   

\section{The Dicke model with a single Bosonic mode}

We now use the formalism described in the previous section to consider the information geometry of 
the Dicke model (see, e.g \cite{pengli},\cite{leonardi}, \cite{vidal1}), which is of great relevance in quantum optics. 
This model describes a bosonic mode interacting with a system of $N$ two-state atoms, via
a dipole interaction, and the Hamiltonian is given by (we will set $\hbar = 1$ in what follows) \cite{emary}
\begin{equation}
H = \omega_0J_z + \omega a^{\dagger}a + \frac{\lambda}{\sqrt{2j}}\left(a^{\dagger} + a\right)\left(J_+ + J_-\right)
\label{dicke}
\end{equation}
Here, $a$ and $a^{\dagger}$ represent the bosonic mode annihilation and creation operators respectively (of frequency $\omega$).
The $J$'s are the collective angular momentum operators obtained by summing over the atomic spin operators, and $\lambda$ is the atom-field coupling 
strength, which drives QPTs in the model, and will be referred to as the driving parameter of the model in sequel. $\omega_0$ is the difference 
between the energies of the two states of the atom. We also set $j = \frac{N}{2}$, where $N$ originates from the volume factor of the cavity in the original 
Dicke Hamiltonian \cite{pengli}. 
In \cite{emary}, it was shown that this choice of $j$ describes the $N$-atom two-level system by a single $\left(N+1\right)$-level system.

In order to understand the information geometry of the Dicke model, we 
treat {\it both} $\omega$ and $\lambda$ as geometric parameters. Effectively, we consider a family of Hamiltonians of the form in eq.(\ref{dicke}), which 
smoothly depends on $\omega$ and $\lambda$. From a geometric perspective, it is natural that all the parameters of the model are treated at par, 
and we are simply looking at a class of models with non-zero values of de-tuning. 

We will first consider the Dicke model in the TDL, using the rotating wave approximation (RWA). The Hamiltonian is given by
\begin{equation}
H_{\rm RWA}=\omega_0J_z+\omega a^\dagger a+\frac{\lambda}{\sqrt{2j}}(a^\dagger J_- +aJ_+)
\label{dickerwa}
\end{equation}
To diagonalize this Hamiltonian in the normal phase\footnote{The presence of extra terms in the Hamiltonian for the super-radiant phase
in the RWA makes it difficult to diagonalize it using position and momentum
representation, and we will confine ourselves to the normal phase in this analysis.}
we follow the method of \cite{emary}. First, we record the standard Holstein-Primakoff representation of the angular momentum operators 
\begin{equation}
J_+ = {b^\dagger}\sqrt{2j-{b^\dagger}b};~~  J_- = \sqrt{2j-{b^\dagger}b}~{b};~~J_z={b^\dagger}b-j
\end{equation}
and in the thermodynamic limit $N\rightarrow\infty$, we obtain, using these, 
\begin{equation}
H_{\rm RWA}=\omega_0{b^\dagger}b+\omega{a^\dagger}a+\lambda({a^\dagger}b+a{b^\dagger})-j\omega_0
\label{dickerwahp}
\end{equation}
Using the position and momentum operators 
\begin{eqnarray}
x&=&\frac{1}{\sqrt{2\omega}}(a^\dagger+a);~~ p_x=i\sqrt{\frac{\omega}{2}}(a^\dagger-a)\nonumber\\
y&=&\frac{1}{\sqrt{2\omega_0}}(b^\dagger+b);~~ p_y=i\sqrt{\frac{\omega_0}{2}}(b^\dagger-b)
\end{eqnarray}
the Hamiltonian of eq.(\ref{dickerwahp}) is given by
\begin{equation}
H_{\rm RWA}=\frac{1}{2}\left[\omega^2x^2+ p_x^2+\omega_0^2y^2+ p_y^2+2\lambda\sqrt{\omega\omega_0}xy+\frac{2{\lambda}p_xp_y}
{\sqrt{\omega\omega_0}}-\omega-\omega_0\right]-j\omega_0
\label{dickerwafinal}
\end{equation}
Now, we rotate the coordinate and momentum axes by 
\begin{equation}
\left[ \begin{array}{c} {\vec R} \\ {\vec P} \end{array} \right] = \begin{bmatrix} {\mathcal I_1} & 0 \\ 0 & {\mathcal I_2}
\end{bmatrix} \left[ \begin{array}{c} {\vec q} \\ {\vec p} \end{array} \right]
\end{equation}
where $0$ is the $2\times 2$ null matrix, and ${\vec R} = \left(x,y\right)^T$, ${\vec P} = \left(p_x, p_y\right)^T$, ${\vec q} = \left(q_1,q_2\right)^T$, 
${\vec p} = \left(p_1,p_2\right)^T$. ${\mathcal I_1}$ and ${\mathcal I_2}$ are $2\times 2$ matrices given by
\begin{equation}
{\mathcal I_1} = \begin{pmatrix}
{\rm Cos}\gamma & {\rm Sin}\gamma \\ -{\rm Sin} \gamma & {\rm Cos}\gamma \end{pmatrix};~~
{\mathcal I_2} = \begin{pmatrix}
{\rm Cos}\gamma' & {\rm Sin}\gamma' \\ -{\rm Sin} \gamma' & {\rm Cos}\gamma' \end{pmatrix}
\end{equation}
Diagonalizing the Hamiltonian of eq.(\ref{dickerwafinal}) then 
sets ${\rm Tan}2\gamma = \frac{2\lambda\sqrt{\omega\omega_0}}{\omega_0^2 - \omega^2}$ and $\gamma' = \frac{\pi}{4}$, and we have two
uncoupled oscillators :
\begin{equation}
H_{\rm RWA}=\frac{1}{2}\left[\varepsilon_-^2q_1^2+p_1^2\left(1-\frac{\lambda}{\sqrt{\omega\omega_0}}\right)+
{\varepsilon_+^2}q_2^2+p_2^2\left(1+\frac{\lambda}{\sqrt{\omega\omega_0}}\right)\right]-\frac{1}{2}\left(\omega+\omega_0\right)-j\omega_0
\label{rwa1}
\end{equation}
where we have defined the collective excitation modes as $\epsilon_{\pm}$ (denoted as the atomic and photonic branches in \cite{emary}
due to their values at $\lambda=0$) :
\begin{equation}
{\varepsilon_\pm}^2=\frac{1}{2}\left(\omega_0^2+\omega^2\pm\sqrt{(\omega_0^2-\omega^2)^2+4\lambda^2\omega\omega_0}\right)
\end{equation}
The Hamiltonian of eq.(\ref{rwa1}) is then re-quantized with two new bosonic modes, 
\begin{eqnarray}
q_1 &=& \frac{1}{\sqrt{2\epsilon_-'}}\left(c_1^\dagger + c_1\right),~~~p_1 = i\sqrt{\frac{\epsilon_-'}{2}}\left(c_1^\dagger - c_1\right)\nonumber\\
q_2 &=& \frac{1}{\sqrt{2\epsilon_+'}}\left(c_2^\dagger + c_2\right),~~~p_2 = i\sqrt{\frac{\epsilon_+'}{2}}\left(c_2^\dagger - c_2\right)
\end{eqnarray}
with its final form being
\begin{equation}
H_{\rm RWA} = m_-\epsilon_-'\left(c_1^\dagger c_1 + \frac{1}{2}\right) + m_+\epsilon_+'\left(c_2^\dagger c_2 + \frac{1}{2}\right)  - \frac{1}{2}\left(\omega_0 + \omega\right)
- j\omega_0
\end{equation}
where $m_{\pm} = 1 \pm \lambda/\sqrt{\omega\omega_0}$, and $\epsilon_{\pm}' = \epsilon_{\pm}/\sqrt{m_{\pm}}$.
The ground state wave function for the model can now be obtained as 
\begin{equation}
\psi_0(x,y)=\left(\frac{\sqrt{1-\lambda^2/(\omega\omega_0)}\varepsilon_+\varepsilon_-}{\pi^2}\right)^{1/4}e^{-(1/2)\langle{\vec R},A{\vec R}\rangle}
\end{equation}
and the overlap between two infinitesimally separated ground states is 
\begin{equation}
\langle\psi_0|\psi_0'\rangle=2\frac{\left[{\rm Det}A~{\rm Det}A'\right]^{1\over 4}}{\left[{\rm Det}\left(A + A'\right)\right]^{1\over 2}}
\label{overlapdicke}
\end{equation}
where $A=U^{-1}MU$, with 
\begin{equation}
M={\rm diag}\left(\epsilon_-\sqrt{m_-},\epsilon_+\sqrt{m_+}\right);~~
U=\left(\begin{array}{cc} {\rm Cos}\gamma & {\rm Sin}\gamma \\ -{\rm Sin}\gamma & {\rm Cos}\gamma \end{array} \right)
\end{equation}
and $A'$ is obtained by infinitesimally shifting $\lambda\to\lambda'$ and $\omega\to\omega'$. 
\begin{figure}[t!]
\begin{minipage}[b]{0.5\linewidth}
\centering
\includegraphics[width=2.8in,height=2.3in]{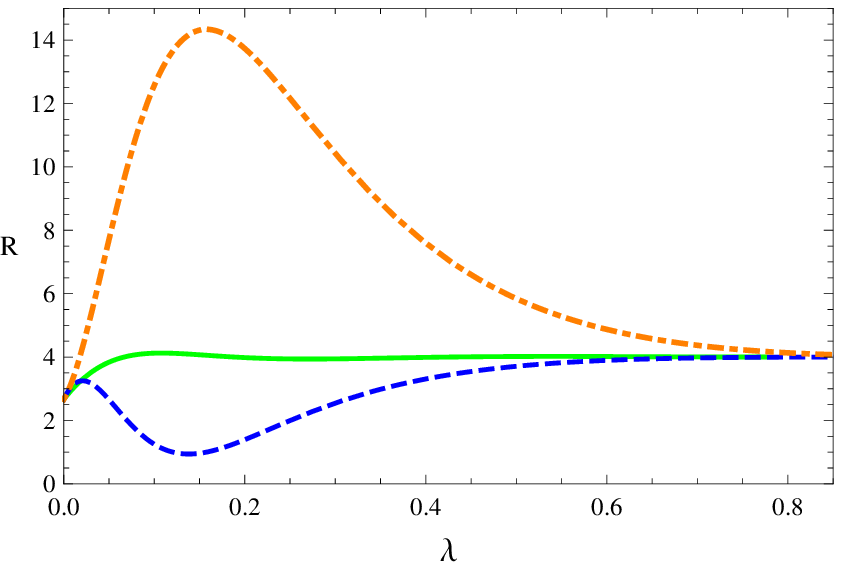}
\caption{(Color online) The scalar curvature $R$ of the Dicke model in the RWA with $\omega_0=1$, plotted as a function of the driving parameter 
$\lambda$ in the normal phase, for $\omega = 0.6$ (solid green), $\omega = 0.8$ (dashed blue) and $\omega = 1.2$ (dot-dashed orange).}
\label{dickerwaR}
\end{minipage}
\hspace{0.2cm}
\begin{minipage}[b]{0.5\linewidth}
\centering
\includegraphics[width=2.8in,height=2.3in]{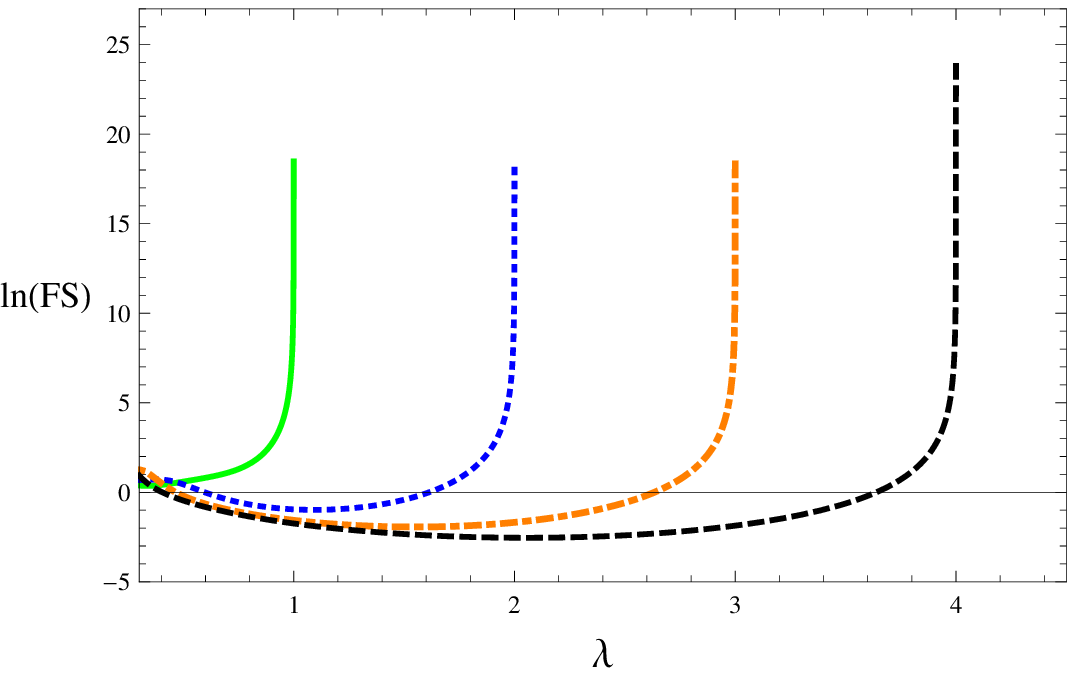}
\caption{(Color online) Log of fidelity susceptibility as a function of $\lambda$ in the RWA
for the Dicke model in the normal phase with $\omega_0=1$, for the paths $\omega = \eta\lambda$, with $\eta=1$ (solid green), $2$(dotted blue),
$3$ (dot-dashed orange) and $4$ (dashed black). }
\label{dickerwafs}
\end{minipage}
\end{figure}

Given the overlap of the ground state wave function of eq.(\ref{overlapdicke}), one can calculate the components of the metric induced on the parameter manifold 
by using eq.(\ref{metric}). The reader will notice that obtaining analytical results for the same is difficult, and we
will present numerical results. First, note that in this case, there is a quantum phase transition at $\lambda_c = \sqrt{\omega\omega_0}$ \cite{emary} where
the excitation energy of the photonic branch, $\epsilon_-=0$. 
Setting $\omega_0=1$ for convenience, we find that all components of the metric tensor diverge at $\lambda_c = \sqrt{\omega}$,
as expected. We point out that the diagonal components of the metric tensor diverge to $\infty$, whereas the off-diagonal 
components diverge to $-\infty$, and the determinant of the metric tensor diverges to positive infinity. 
However, we find that there is no divergence of the scalar curvature at the phase boundary. This indicates that 
the underlying manifold in the parameter space does not exhibit any special behavior here, and we conclude that the divergences of the components of the metric
tensor are artifacts of the coordinate choice, as was the case of the transverse XY model of the previous section (along the Ising transition line). We find that
the scalar curvature $R \to 4$ at the phase boundary for all values of $\omega$. We also calculate the fidelity susceptibility along
the paths parametrized by $\omega = \eta \lambda$ for $\eta = 1,\cdots 4$ using the methods of \cite{gu},\cite{amit1}. We
find that, being a sum of the metric tensor components, this diverges at the phase boundary as expected, 
although this divergence does not have any bearing on the global structure of the parameter manifold. 
Our results are shown in figs.(\ref{dickerwaR}) and (\ref{dickerwafs}), where we have shown numerical plots for the scalar curvature $R$ and the 
log of the fidelity susceptibility, as a function of the driving parameter $\lambda$, for different values of $\omega$. 
We note that the peaks in the scalar curvature in fig.(\ref{dickerwaR}) are possible
mathematical artifacts and do not carry any physical significance. They are absent when we consider the theory without the RWA, to which we presently turn to.
\begin{figure}[t!]
\begin{minipage}[b]{0.5\linewidth}
\centering
\includegraphics[width=2.8in,height=2.3in]{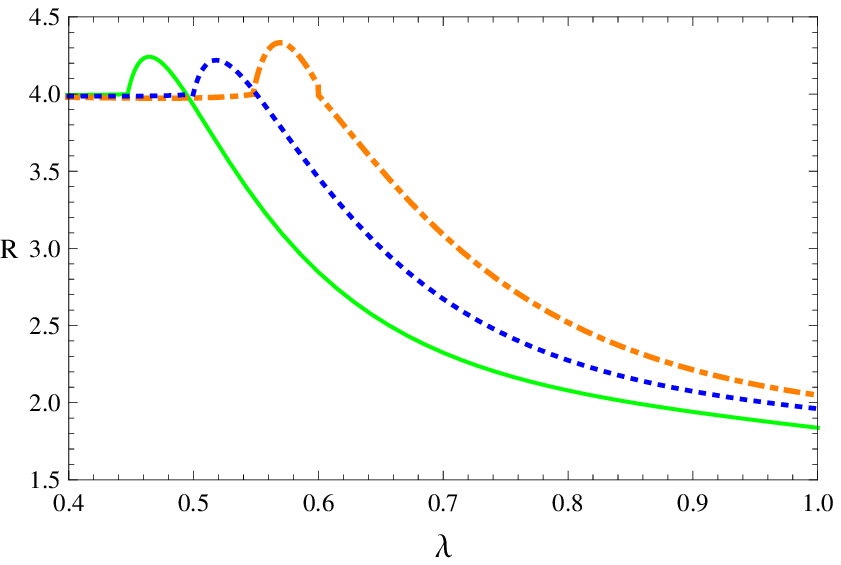}
\caption{(Color online) The scalar curvature $R$ of the Dicke model without the RWA, with ${\tilde\omega_0}=1$, 
plotted as a function of ${\tilde\lambda}$, for ${\tilde \omega} = 0.8$
(solid green), ${\tilde \omega} = 0.999$ (dotted blue) and ${\tilde \omega} = 1.2$ (dot-dashed orange).}
\label{dickeemaryR}
\end{minipage}
\hspace{0.2cm}
\begin{minipage}[b]{0.5\linewidth}
\centering
\includegraphics[width=2.8in,height=2.3in]{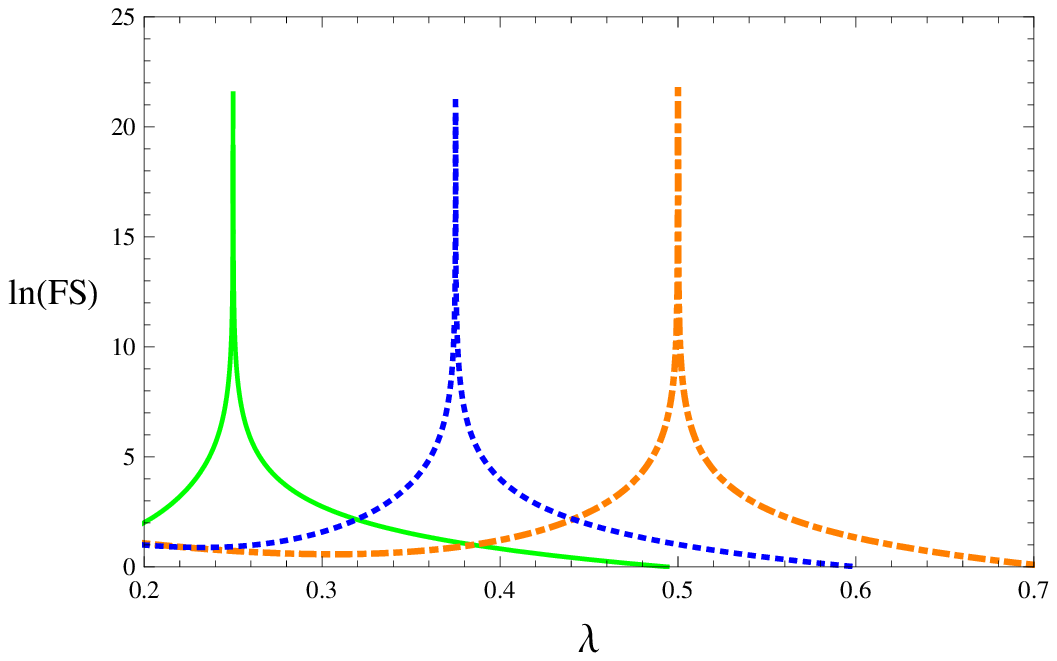}
\caption{(Color online) Log of the Fidelity Susceptibility for the Dicke model without the RWA, for ${\tilde\omega}_0=1$, plotted as a function of 
${\tilde\lambda}$, for the paths 
${\tilde\omega}={\tilde\lambda}$ (solid green), ${\tilde\omega} = 1.5{\tilde\lambda}$ (dotted blue), ${\tilde\omega} = 2{\tilde\lambda}$ (dot-dashed orange).}
\label{dickeemaryfs}
\end{minipage}
\end{figure}

For the Dicke model in the TDL, without the RWA, the results are qualitatively similar. Here, the overlap between two infinitesimally separated ground states in the normal phase 
\footnote{All the quantities are denoted by a tilde here compared to the analysis with the RWA, to avoid confusion.} is given by \cite{zan2},\cite{emary}, 
\begin{equation}
\langle{\tilde\psi}_0\left({\tilde\omega},{\tilde\lambda}\right)|{\tilde\psi}_0
\left({\tilde\omega}',{\tilde\lambda}'\right)\rangle = 2\frac{\left[{\rm Det}{\tilde A}~{\rm Det}{\tilde A}'\right]^{1 \over 4}}
{\left[{\rm Det}\left({\tilde A} + {\tilde A}'\right)\right]^{1\over 2}}
\label{overlapdicke1}
\end{equation}
where the matrix ${\tilde A}$ is 
\begin{equation}
{\tilde A} = \frac{1}{2}\begin{pmatrix}
\alpha - \beta{\rm Cos}2{\tilde\gamma} & \beta{\rm Sin}2{\tilde\gamma}\\
\beta{\rm Sin}2{\tilde\gamma} & \alpha + \beta{\rm Cos}2{\tilde\gamma}
\end{pmatrix}
\end{equation}
with $\alpha,\beta = {\tilde\epsilon}_+\pm {\tilde\epsilon}_-$, ${\tilde\gamma} = \frac{1}{2} {\rm Tan} ^{-1}\left(\frac{4 {\tilde\lambda}\sqrt{{\tilde\omega}{\tilde\omega}_0}}
{{\tilde\omega}_0^2-{\tilde\omega}^2}\right)$. The
matrix ${\tilde A}'$ is obtained by replacing ${\tilde\omega} \to {\tilde\omega}'$ and ${\tilde\lambda}\to {\tilde\lambda}'$ in the matrix ${\tilde A}$, and similarly 
for ${\tilde\epsilon}_{\pm}$.
Here, ${\tilde\epsilon}_{\pm}$ are the energies of the two uncoupled oscillator
modes (atomic and photonic branches), given by
\begin{equation}
{\tilde \epsilon}_{\pm}^2 = \frac{1}{2}\left({\tilde\omega}_0^2 + {\tilde\omega}^2 \pm \sqrt{\left(
{\tilde\omega}^2 - {\tilde\omega}_0^2\right)^2 + 16{\tilde\lambda}^2{\tilde\omega}{\tilde\omega}_0}\right)
\label{epm}
\end{equation}
At ${\tilde\lambda}_c = \frac{1}{2}\sqrt{{\tilde\omega}{\tilde\omega}_0}$, ${\tilde\epsilon}_-=0$, and the 
system undergoes a QPT. 

For the overlap function of eq.(\ref{overlapdicke1}), one can calculate the metric components and the scalar curvature just as before, by treating ${\tilde\omega}$ and 
${\tilde\lambda}$ as geometric parameters. Again we find that the diagonal (off-diagonal) components of the metric tensor diverge to $\infty$ ($-\infty$)
and the determinant of the metric diverges to $\infty$.
A similar analysis can be carried out in the super-radiant phase using its ground state wave function derived in \cite{emary}, and the 
scalar curvature $R$ can again be obtained as a function of ${\tilde\lambda}$ and ${\tilde\omega}$ (we set ${\tilde\omega_0}=1$). Combining this with
the analysis for the normal phase, we obtain $R$ in the entire region of the parameter space. 
In fig.(\ref{dickeemaryR}), we plot $R$, as a function of the driving parameter ${\tilde\lambda}$ for 
various values of ${\tilde\omega}$. In the normal phase, the scalar curvature does not show any peak (present with the RWA) and its value remains
$\simeq 4$ throughout, consistent with what we obtained with the RWA. 
In the super-radiant phase, $R \to 4$ at the phase boundary, and shows a maximum, slightly away from the boundary, but we are not aware of 
a physical reason for the same. We also find that in the limit of large ${\tilde \lambda}$, $R \to 0$, which is consistent with the fact that in this limit,
the modes ${\tilde \epsilon}_+$ and ${\tilde \epsilon}_-$ decouple \cite{emary}, and the ground state wave-function is independent of $\lambda$. As a
result of this, the parameter manifold becomes one-dimensional in this limit, and hence has zero curvature. $R$ being a coordinate independent
quantity, our conclusion here is that the information geometric manifold is smooth at the location of the 
QPT. The scalar curvature is also seen to be continuous across the transition, from fig.(\ref{dickeemaryR}), in contrast to the result for the 
transverse XY model. Finally, fig.(\ref{dickeemaryfs})
shows the results for the log of the fidelity susceptibility, which expectedly diverges at the phase transition. 

\section{Discussions and Conclusions}

In this concluding section, for the sake of completeness, let us first mention some cases that we believe, deserve further attention and analysis. Our first example is that of the 
Lipkin-Meshkov-Glick model for $N$ spins (see e.g \cite{vidal2} for a finite temperature analysis, and references therein), given by the Hamiltonian 
\begin{equation}
H_{\rm LMG}=-\frac{1}{N}\sum_{(i<j)}(\sigma_{i}^{x}\sigma_{j}^{x}+\gamma\sigma_{i}^{y}\sigma_{j}^{y})-h\sum_{i}\sigma_{i}^{z}
\end{equation}
where the $\sigma$'s are the Pauli matrices, and $\gamma$ and $h$ are the anisotropy parameter and the magnetic field respectively. 
After diagonalizing the Hamiltonian using Holstein-Primakoff and Bogoliubov transformations (we omit the details of the standard calculations here), 
it can be shown that the induced metric on the parameter manifold has the components 
\begin{equation}
g_{hh} = \frac{(\gamma -1)^2}{32 (h -1)^2 (\gamma -h)^2};~
g_{\gamma\gamma} = \frac{1}{32 (\gamma - h)^2};~
g_{h\gamma} = -\frac{\gamma -1}{32 (h -1) (\gamma -h)^2}
\end{equation}
It can be seen that the determinant of the metric is zero, signaling that information geometry is ill defined in this model. A similar conclusion has been reached
by different methods in \cite{scherer}, and this needs further investigation. 
In the context of the Kitaev honeycomb model, we find that the parameter manifold is smooth, i.e
does not show any divergence. The scalar curvature here shows the same qualitative behavior as the fidelity susceptibility \cite{gukitaev}, i.e $R$ 
show a series of peaks in the gapless region, exactly coinciding with the peaks of the fidelity susceptibility \cite{gukitaev}. 
Information geometry of this model also needs to be understood further. 

To summarize, in this paper, we have undertaken a critical analysis of information geometry in quantum phase transitions, in the context of the Dicke model of quantum optics
in the thermodynamic limit. By analyzing the scalar curvature of the parameter manifold, which relates to coordinate independent properties of the same, 
we have shown that this manifold is smooth for the Dicke model, and does not show any singular behavior at phase transitions,
although the components of the metric tensor diverge there.  We have studied the geometry for the Dicke model, both with and without the rotating wave approximation,
and found consistent results for the scalar curvature of the parameter manifold. We have also shown that the scalar curvature is continuous across the phase boundary. 
Our results indicate that purely from a geometric perspective, the divergences in the metric tensor in this model can be removed by appropriate coordinate transformations. 

Our results also point to the fact that information geometry does not seem to provide a universal characterization of second order QPTs via the scalar
curvature of the parameter manifold. Apart from the Dicke model, where $R$ is regular, 
we have demonstrated this explicitly in the context of the XY spin chain, where it was shown 
that $R$, in general, does not have a simple relation with the correlation length at a QPT. 
These statements, along with our discussion in the beginning
of this section on the LMG and the Kitaev honeycomb models, indicate that for QPTs, the geometry of the parameter manifold needs to be studied 
on a case by case basis. 

This should be compared to the case of classical phase transitions, in which, in all known examples, the scalar curvature of the information metric diverges
at second order critical points, indicating a genuine (irremovable) singularity of the parameter manifold. This universal behavior can be related to the fact that near the critical 
points for CPTs, $R$ is conjectured to be proportional to the correlation volume, which is not the case for QPTs, as we 
have stated. In fact, for CPTs, some components of the metric tensor 
are known to vanish at criticality (so that the geometry becomes ill defined, i.e, singular), but for QPTs, this is not the case. Exploring this direction further might be an 
useful exercise to undertake, particularly in the context of some known models in which classical analysis in the thermodynamic limit pinpoints the QPT 
as well \cite{diptiman},\cite{amit}. Finally, in the absence of any singularity of the parameter manifold, it is interesting to ask how information geometry distinguishes between
different phases of a theory that admits a QPT, in a coordinate independent manner. This issue will be addressed in a forthcoming publication.   

\begin{center}
{\bf Acknowledgements}
\end{center}
It is a pleasure to thank Amit Dutta and Harshawardhan Wanare for many illuminating discussions, and Diptiman Sen for helpful correspondence.
The work of SM is supported by grant no. 09/092(0792)-2011-EMR-1 from CSIR, India.\\

\end{document}